\documentclass[12pt,preprint]{aastex}

\shorttitle{Southern close binary stars~II}
\shortauthors{Rucinski \& Duerbeck}

\begin{document}

\title{Radial Velocity Studies of Southern Close Binary
Stars.~II. Spring/Summer 
Systems\footnote{Based on data obtained at the European
Southern Observatory.}}

\author{Hilmar W. Duerbeck}
\affil{WE/OBSS, Vrije Universiteit Brussel, Pleinlaan 2,
B-1150 Brussels, Belgium}
\email{hduerbec@vub.ac.be}

\and

\author{Slavek M. Rucinski}
\affil{David Dunlap Observatory, University of Toronto \\
P.O.~Box 360, Richmond Hill, Ontario, Canada L4C~4Y6}
\email{rucinski@astro.utoronto.ca}

\begin{abstract}
Radial-velocity measurements and sine-curve fits to the orbital
velocity variations are presented for 14 close binary stars,
S~Ant, TT~Cet, TW~Cet, AA~Cet, RW~Dor, UX~Eri, YY~Eri, BV~Eri,
CT~Eri, SZ~Hor, AD~Phe, TY~Pup, HI~Pup and TZ~Pyx. All are
double-lined binaries and all except the last one are
contact binaries. The orbital data must be considered
preliminary because of the relatively small number of observations
(6 -- 12), a circumstance which is partly compensated by the
good definition of the broadening functions used
for the RV determinations.
\end{abstract}


\section{INTRODUCTION}
\label{sec1}

This is the second paper of a two-paper 
series of radial velocity (RV)
observations of close binary systems conducted at the European
Southern Observatory in 1996 and 1998. The first paper
\citep[ Paper~I]{RD06} presented data obtained on 4 nights in
August 1998 for 17 targets, a mixture of
contact binaries possibly offering reasonable orbital
solutions with a selection of variables suspected to be
contact binaries. The current paper is very similar in
spirit: It gives radial velocity data and preliminary
orbital solutions for 14 similar targets observed on 5 nights
of November 28 to December 2, 1996, in the later
part of the southern sky.

The goals of this paper, similarly to those of 
Paper~I, are close to the series of radial velocity
studies currently conducted at the David Dunlap 
Observatory (DDO). This series has recently reached, 
with the eleventh paper \citep{ddo11} (where references 
to the previous papers and many common details can be found), 
the round number of 100 well covered RV orbits.
The instrumentation and data analysis techniques, are 
explained in Paper~I. The observations were 
carried with the ESO La Silla 1.52-m telescope
and a B\&Ch Cassegrain spectrograph. 
The broadening functions were extracted from the 
wavelength region of  401.6 -- 499.8 nm. 
The spectra have a BF resolution of 
typically $\simeq 23 - 27$ km~s$^{-1}$.
Stellar exposure times ranged between 10 and 20 min, 
depending on brightness; each stellar exposure 
was followed by a He-Ar spectrum. 
Spectrum extraction and wavelength calibration was
carried out with ESO MIDAS software 
system\footnote{http://www.eso.org/projects/esomidas}. 
As a comparison star template, we used 68 Eri (F5 V), 
for which a radial velocity of
$+10.1~\rm km~s^{-1}$ was adopted~\citep{nord04}.

In terms of the presentation, we again stay close to the format
of Paper~I, the only difference being that the targets are
discussed simply in the constellation order in 
Section~\ref{sec2}. We see no obvious
cases of pulsating stars in this group of the
targets; only CU~Eri A and CU~Eri~B do not appear to be
close binaries. 
We describe our results in the context of existing
photometric data from the literature and the Hipparcos
project. We also utilize the mean $(B-V)$ color indexes 
taken from the Tycho-2 catalog \citep{tycho2}
and the photometric estimates of the spectral types
using the relations published by \citet{bessell79}.
Spectral types are taken uniformly from the 5 volumes
of the Michigan Catalogue of HD Stars
\citep{houketal75,houk78,houk82,houketal88,houketal99};  
hereafter quoted as HDH.
Because of the high incidence of companions to 
contact binary stars
\citep{priruc2006}, we checked all stars for possible 
membership in visual systems using the Washington Double Star 
Catalog (WDS)\footnote{http://ad.usno.navy.mil/wds/}.

Figures~\ref{fig1} and \ref{fig2}
present the broadening functions for all
targets at phases selected for best visibility of both
components. Figures~\ref{fig3} and \ref{fig4}
show the preliminary radial
velocity orbital solutions. The RV data are given in
Table~\ref{tab1}, while parameters of the orbital solutions are
listed in Table~\ref{tab2}. For most stars we used the primary
eclipse predictions from the Cracow on-line database
\citep{kreiner2001,kreiner2004}\footnote{http://www.as.ap.krakow.pl/ephem/,
version May 2006.}. These predictions can be restored
from the observed times of conjunctions ($T_0$) and the
observed {\it minus\/} predicted deviations $O-C$, as
given in Table~\ref{tab2}.

\section{INDIVIDUAL BINARY SYSTEMS}
\label{sec2}

\subsection{S Ant}

S Ant (HIP~46810, HD~82610) with $V_{\rm max}=6.29$ is 
one the brightest contact binaries (this and the following
estimates of $V_{\rm max}$ were obtained from the
Hipparcos magnitudes \citep{hip} applying color dependent
corrections $H_p - V$). As its name implies, it is one 
of the first variable stars recognized as such in 
the southern hemisphere.

S Ant has been a frequent target of photometric
observations, but the only previous spectroscopic 
observations were by \citet{popper56} who saw only 
one component with $V_0=-1.2 \pm 1.0$
km~s$^{-1}$ and $K_1=92.3 \pm 1.5$ km~s$^{-1}$. 
In contrast, we could easily
detect the secondary component in the broadening 
functions (BFs), as shown in Figure~\ref{fig1}. 
The center-of-mass velocity is very different from
that observed previously, as are both semi-amplitudes $K_i$
(Table~\ref{tab2}). 
\citet{russo82} encountered some difficulties with
producing a contact model for the binary suggesting possible
problems with Popper's spectroscopic data.
They derived the mass ratio $q_{\rm ph}=0.59 \pm 0.02$,
which differs substantially from our 
$q_{\rm sp}=0.33 \pm 0.02$. This is not surprising 
for a partially eclipsing system with a featureless light curve,
for which a photometrically derived mass ratio 
is notoriously unreliable. We note that
our bootstrap errors of $K_i$ are quite large,
primarily due to the small number -- rather than the low accuracy --
of individual observations; these uncertainty estimates are
more realistic than the formal least-squares errors.

The moment of the minimum as given in Table~\ref{tab2}
agrees perfectly well with the Cracow ephemeride
given in \citet{kreiner2001} on the assumption 
that the trend indicated by the $O-C$
diagram has continued and reached $+0.060$ day
at the time of our observations. This value
is uncertain and requires confirmation by
photometric observations.

Judging by the light curve (e.g.\ from Hipparcos),
the current radial velocity data and the shapes of the BFs,
S~Ant is a very typical A-type contact binary. 
Its spectral type is F3V (HDH) which agrees with 
$B-V=0.30$ from \citet{tycho2}.
Its orbit almost certainly does not have any eccentricity, so its
inclusion among early-type stars with potentially $e \ne 0$
\citep{abt05} and the spectral type A6-F0IV-V are based on obsolete
material.

\subsection{TT Cet}

TT Cet (HIP 8294) previously had not been observed spectroscopically,
probably because of the relative faintness of the system at
$V_{\rm max}=10.83$. The light curve
suggests components of very different temperatures giving TT~Cet
a classification of EB. The star appears in the compilation of
\citet{shaw94} as a ``near-contact'' binary. Indeed, the secondary
is a faint and cool star because its signature is
just barely present in our BFs (Figure~\ref{fig1}).
However, the SB2 orbit is relatively well defined although we have
only the second half of the orbit properly covered by observations
(Figure~\ref{fig3}). 
The binary appears to be a semi-detached or more likely
a detached binary with a very faint secondary component.
The mass ratio is $q_{\rm sp}=0.39 \pm 0.07$. The spectral type is 
unknown. From $B-V=0.40$ derived from Tycho-2 observations,
one can estimate the spectral type of the 
primary star to be near F4V.

\subsection{TW Cet}

TW Cet (HIP 8447)
is another relatively frequently photometrically observed
binary with insufficient spectroscopic data. At $V_{\rm max}=10.32$
and with a short period 0.32 days, it requires a moderately
large telescope to avoid RV phase smearing.
\citet{struve50} attempted to obtain radial velocities and
obtained a preliminary orbit with $V_0=+20$ km~s$^{-1}$,
$K_1=135$ km~s$^{-1}$ and $K_2=255$ km~s$^{-1}$. They stressed
that the two components appeared to be equal in strength.
Our BFs indeed show very similar components, but
the orbital semi-amplitudes are different, 
resulting in a mass ratio $q=0.75 \pm 0.03$.

With its mass ratio quite close to unity, $B-V=0.60$
(Tycho-2) implying a spectral type of about G1V, and very
broad signatures of both components in the broadening functions
(Figure~\ref{fig1}), TW~Cet appears to be a genuine, late-type
contact binary. The binary is a component of the visual system
WDS~01489-2053. The fainter companion 
($\Delta m \simeq 2$) slowly
changed in angular separation from 8.3 arcsec in
1923 to 8.8 arcsec in 1998. It was not included in 
the spectrograph slit. 

\subsection{AA Cet}

AA Cet (HIP 9258, HD 12180)
is a relatively bright contact binary ($V_{\rm max} = 6.58$)
with unequally deep eclipses giving it a light curve
classification of EB.
It forms part of the visual binary WDC~01590-2255 and 
has a slightly fainter ($\Delta m \simeq 0.3$) 
companion at a separation of 8.6 arcsec 
which was not included in the spectrograph
slit. We obtained, however, one spectrum of the companion
giving a radial velocity of $29.6 \pm 1.2$ km~s$^{-1}$. This
is marginally consistent with
the center of mass velocity of the binary, $V_0=32.9 \pm 2.1$
km~s$^{-1}$.

The radial velocity orbit of AA~Cet
is well defined. The star appears to be a rather typical
contact binary with a mass ratio of $q_{\rm sp}=0.35 \pm 0.02$.
The spectral type in HDH, A7/8V + G, might reflect the combined
contribution of both visual components. However, $B-V=0.38$
(Tycho-2) corresponds to F3V rather than A7/8V, so that the
spectral classification may have suffered from
contamination by the visual companion.
The color measurement $b-y=0.256$ \citep{WK83} also
suggests F3V.
\citet{chamb81} gave the spectral type of the binary as F2
and noted that the companion, with a spectral type of F5,
shows sharp, double-lined spectra, indicating that
itself it is a close binary system. Our single spectrum of
the companion does not show any doubling of the lines 
while the BF is very sharp as expected for a slowly rotating
single star.

\subsection{RW Dor}

RW Dor (HIP 24763, HD 269320) has been known as an
variable since its discovery by Leavitt in 1906; but it was
\citet{hertz28} who classified it as a W~UMa system.
It has been a subject of
numerous photometric studies. The only spectroscopic
radial velocity orbit is by \citet{hild92} who found
$V_0= 66.5 \pm 4.8$ km~s$^{-1}$, $K_1=130.5 \pm 5.7$
km~s$^{-1}$ and $K_2=191.5 \pm 3.1$ km~s$^{-1}$, thus
implying $q_{\rm sp}=0.68 \pm 0.04$. Our orbit is
different in that $V_0$ is by 25 km~s$^{-1}$ smaller and
both semi-amplitudes $K_i$ are larger; the mass ratio is,
however, similar: $q_{\rm sp}=0.63 \pm 0.03$. An under-estimation
of the semi-amplitudes is a typical problem of
insufficient resolution which may have been the result of
the use of the cross-correlation function (CCF), in
contrast to our much better resolving BF technique.
On the other hand, our phase coverage is very sparse, with
only 3 observations not falling within the conjunctions.

\citet{hild92} assumed a spectral type of K1V following
\citet{mart89}. $B-V=0.66$ (Tycho-2) suggests an 
earlier spectral type of about G4/5V with $V_{\rm max}=10.97$.
The binary is a contact
system of the W-type, i.e.\ the less massive, but brighter
component is eclipsed at the primary eclipse.

\subsection{UX Eri}

The contact binary
UX Eri (HIP 14699) was a subject of a recent DDO study
\citep[ DDO-3]{ddo3} to which the reader is referred to for more
details. Four of the current ESO observations were used in that
study in addition to 24 DDO observations. Thus, the results
of DDO-3 are certainly more reliable and should be used in
the future. The current data are included here only for
completeness. A separate orbital solution based on only these
seven points (Table~\ref{tab2}) 
gives smaller values of both $K_i$,
which can be a result of both a lower spectral resolution and
of a much smaller number of observations.

\subsection{YY Eri}

YY Eri (HIP~19610, HD~26609) is a bright
($V_{\rm max} \simeq 8.24$) contact system, a frequent subject
of photometric investigations.
A spectroscopic study was presented by
\citet{struve47}, with small corrections by \citet{huruhata53},
giving the two semi-amplitudes, $K_1=130$ km~s$^{-1}$
and $K_2=220$ km~s$^{-1}$ and thus $q_{\rm sp}=0.59$.
A light curve synthesis solution was presented by
\citet{macer82} where the mass ratio was kept at the spectroscopic
value. Later, \citet{nesci86} presented orbital
spectroscopic data with $V_0=-15$ km~s$^{-1}$, $K_1=104$
km~s$^{-1}$ and $K_2=271$ km~s$^{-1}$; with the
proximity effects included, $q_{\rm sp}=0.401$.

Our new orbit is well defined, agreeing moderately
well with that of \citet{nesci86}. Our new value of
$q_{\rm sp}=0.44 \pm 0.01$ does not include the
proximity effects. The component of the larger mass
shows better defined and stronger
peaks in the BFs, but this component
is in front during the primary (deeper) eclipse,
so the binary is a W-type contact system.
The spectral type of HDH is G3/K0+(F/G)
indicating a complex spectrum. We have no indications of
the multiplicity of the system, neither from the BFs, nor
from the visual star catalogs. $B-V=0.64$ suggests a
spectral type about G3V.

\subsection{BV Eri}

BV Eri (HIP~18080, HD~24327) is a well known, bright
($V_{\rm max}=8.13$) close binary
with unequal surface brightness of the components giving it
a light curve classification of EB. \citet{gu99}, following
several investigations of similar systems by Ka{\l}u\.{z}ny
(e.g.\  \citet{kal86} where previous analyses are cited) considers
the system to be a contact one, but with a very large
difference in the brightness of both components.
HDH gave a spectral type F2V, while the mean Tycho-2
color, $B-V=0.40$ would suggest a slightly later type of F3/4V,
or some small amount of reddening.

The only previous radial velocity study is by \citet{baade83}
who were able to see only one component giving $V_0=-44.4 \pm 3.4$
km~s$^{-1}$ and $K_1=53.1 \pm 3.6$ km~s$^{-1}$. We can see both
components in our BFs and the orbital solution
is relatively well defined. Our mass ratio, 
$q_{\rm sp}=0.30 \pm 0.02$, is different from 
Gu's $q_{\rm ph}=0.25$, which was given with an
unrealistically small uncertainty of 0.001.

\subsection{CT Eri}

CT Eri (HIP~20943) is a system very similar to BV~Eri, 
except that it is fainter ($V_{\rm max}=9.95$) 
and has a fainter companion ($\Delta m \simeq 1.8$) 
at a separation of 10 arcsec, forming the
double system WDS~04294-3335. The spectral type
is probably also similar: Judging by the mean $B-V=0.38$ (Tycho-2),
the spectral type may be F2/3V. The Simbad database quotes
F0. Its variability was announced by \citet{stroh68}.
No light curve solutions nor radial velocity studied
have been attempted so far.

We see faint signatures of the secondary component of CT~Eri
in our BFs, although they are always difficult to measure. 
The formally derived mass ratio is practically 
identical to that of BV~Eri,
but its uncertainty is very large: 
$q_{\rm sp}=0.30 \pm 0.09$, mostly
because we could detect the faint secondary in only two
observations. The BFs in the second half of the orbit were very
difficult to interpret with asymmetries of 
the primary peak and an absence of the secondary peak.

\subsection{CU Eri}

This visual binary has been included in our observations 
because it was listed in the list of \citet{wood} 
as an eclipsing binary with a period of
0.63 days, but this was based on preliminary information only.
\citet{ss87} noted its variability, but could not establish its type. 
CU~Eri is part of the visual system WDC~02470-1320
with a separation of 3.7 arcsec and the visual magnitudes
of $V=8.74$ and 13.5. Our BFs (Figure~\ref{fig2})
indicate that spectral lines of the component B are slightly wider
yielding $V\sin i \simeq 20$ km~s$^{-1}$ with
possibly some asymmetry, while the component A shows a
sharp, single peak typical for a slowly rotating star observed at
our resolution. The spectral type of A in
HDH is G8/K0III. Neither component appears to be a close binary.

\subsection{SZ Hor}

SZ Hor (HIP 14488) has not been much observed since its discovery
\citep{stroh67a} probably because of its faintness,
$V_{\rm max}=11.01$. Judging by
its Tycho-2 color, $B-V=0.37$, its spectral type is about F3V.
Our orbit is well defined and could be taken to indicate
a typical contact binary of the A-type if not for the
very different depths of the two eclipses giving
it the light curve classification of EB.
The mass ratio is $q_{sp}=0.47 \pm 0.04$.

\subsection{AD Phe}

AD~Phe (HIP 5955) has been recognized as a variable star for many
years \citep{stroh69}. \citet{McFH87} obtained a good light curve
and attempted to solve it making two doubtful (and recognized as such)
assumptions of $q=0.5$ and $q=1.0$. In fact, we find
$q_{\rm sp}=0.37 \pm 0.01$. 
Similarly to YY~Eri, the stronger and better
defined component is the more massive one, but it is the one
eclipsed at the slightly shallower minimum,
so this is a W-type contact binary.
No spectral type is available, but the mean 
$B-V=0.56$ suggests F9V or G0V. The star is moderately 
faint at $V_{\rm max}=11.28$.

\subsection{TY Pup}

TY Pup (HIP~36683, HD~60265) is one of the brighter southern
contact binaries ($V_{\rm max}=8.41$) and, as such,
the subject of many photometric investigations.
The star appears as a variable the first time
in \citet{hertz28}. The light curve suggests
a contact binary possibly undergoing total eclipses.
The only previous spectroscopic data have been
obtained by \citet{struve50a} who saw two periodicities (both
incorrect, 0.58 and 9.7 days) and determined 
$V_0=+28$ km~s$^{-1}$ and $K_1=32$ km~s$^{-1}$. 
He classified the spectrum as A9n.
The spectral classification of HDH is F3V, which agrees very well
with the Tycho-2 mean color index $B-V=0.36$.

The light curve synthesis modeling was attempted by
\citet{macer82} who obtained $q_{\rm ph}=0.32 \pm 0.05$,
and by \citet{gu93}, who obtained a very different and
unrealistically accurate $q_{\rm ph}=0.1846 \pm 0.0020$. Although
we only have few observations, the mass ratio differs
from both estimates, $q_{\rm sp}=0.25 \pm 0.03$. The system is
a very typical A-type contact binary without any obvious
complications.

\subsection{HI Pup}

HI Pup (HIP 36762) is a relatively faint ($V_{\rm max}=10.36$)
contact binary which has not had any photometric or
radial velocity observations. Discovered in 1949, the system
was classified and its light curve constructed from 
photographic observations by \citet{hoff56}. 
\citet{sahade63} suggested that it is a possible member
of the open cluster Cr~173, but this matter has not been
established yet. The Tycho-2 mean $B-V=0.51$ corresponds to
a spectral
type about F6V, but there exist no direct classifications of
the star. A companion with $B-V =1.68$ (approximately M3V)
is located 52 arcsec away \citep{hip}, but the star is not
included as a visual double system in the WDS Catalog.

We have only 5 radial velocity observations for HI~Pup, so
the radial velocity orbit must be considered very preliminary.
The mass ratio is small, $q_{\rm sp}=0.19 \pm 0.06$.

\subsection{TZ Pyx}

TZ Pyx (HIP 42619)
was discovered as a variable star by \citet{stroh66} who
determined the orbital period to
be 0.697 days \citep{stroh67b}. 
Later Hipparcos observations \citep{hip}
showed that the binary has an orbital period of 2.3 days
with a light curve showing relatively short, well defined
eclipses. It is definitely a detached binary with well
separated components, yet
the star still appears as a W~UMa-type binary in the
Simbad database with the spectral type as an uncertain A.
The Tycho-2 photometric data, $B-V=0.27$, suggest a
spectral type A8/9V at $V_{\rm max}=10.68$.

Our radial velocity orbit is the best among this group of
targets, mostly because of sharp signatures in the broadening
functions which were easy to measure for radial velocities.
The mass ratio is
very close to unity, $q_{\rm sp}=0.965 \pm 0.020$, with the
slightly less-massive component (i.e.\ the one
showing the larger semi-amplitude) giving
a marginally stronger signature in the BF's.

\section{SUMMARY}
\label{summ}

The current paper is the second and last of the two papers
describing a short program of radial velocity observations
of southern contact binary stars conducted at the
European Southern observatory in December 1996 and 
August 1998. Paper~I gave the full
background and explanations for the rationale of this
program, and contained results for 17 targets observable
in the August season in 1998.

The current paper presents results for
14 binary stars and two suspected variable stars (both components
of the visual double CU~Eri) which were observed in 1996.
All the observed 14 binaries turned out to be
double-lined (SB2) permitting derivation of preliminary RV orbits
for all of them. Thus, three previously considered as
SB1 systems, S~Ant, BV~Eri and
TY~Pup, have new, complete if sparsely covered RV orbits. Our
broadening function technique permitted to obtain improved
RV orbits for the previously known SB2 systems TW~Cet, RW~Dor
and YY~Eri. We confirm that TZ~Pup is definitely not a contact
binary star, but rather a detached system of almost identical
A-type stars, in full agreement with the results of Hipparcos
which showed that the period is 2.3 days.

We see very faint signatures of secondary components 
in some of our systems (TT~Cet, CT~Eri, BV~Eri, SZ~Hor), 
but in all cases we are confident that these are real 
detections. We could detect them because of the advantageous 
properties of the BF technique; indeed, in most cases the 
secondaries are undetectable using the Cross Correlation 
Function approach. The phase dependencies of the secondary star 
velocities are exactly as expected. However, these results 
will be testable in the future when data with higher spectral 
resolution and $S/N$ become available.

Four systems discussed in this paper and in Paper~I have been 
investigated independently from ESO and -- at a later date 
-- from DDO (the DDO solution for UX~Eri utilized a small
number of ESO observations so it was not fully independent). 
For these systems, the mass-ratios $q$ derived from ESO 
are on the average 6\% larger (with a range of 0\% to 15\%), 
and the $V_0$ center-of-mass velocities from 
ESO by $5.7 \pm 3.1~\rm km~s^{-1}$ (s.d.) more negative. It
should be mentioned that the original goal of the 
current survey was to obtain reasonably accurate 
$V_0$ velocities for the determination of space motions, 
as well as first estimates of the system properties, 
within the allotted amount of observing time.

Concerning the statistics of Southern hemisphere
contact binary stars: The two
parts of the curent spectroscopic 
program resulted in observation of 23 contact
binaries of which 12 are brighter than 
$V_{\rm max} < 10$. Taking into
account that only about 1/2 of the southern sky was
observed, we see that the numbers are small, much smaller
than for the northern hemisphere. This dearth of data
for the southern hemisphere was pointed by 
\citet{priruc2006} and is in a great need for rectification.
Within the strict limit of $V_{\rm max} = 10$, 
there are currently 20 southern contact systems with 
reliable radial velocity curves versus 58 
similar northern systems. 
We note that the inequality between the hemispheres 
manifests itself not only in the lack of radial velocity data, 
but also in the number of detected contact binaries:
To the same magnitude limit, we know 
56 southern systems versus 72 northern systems, as based on 
the catalogue of \citet{pri03}). On the basis of the apparent
relative frequency among FGK dwarfs of 0.2\%, derived by
\citet{due84} and fully confirmed later \citep{rci02,rci06},
one can expect about 800 to 1000 contact binaries on the whole
sky brighter than $V_{\rm max}=10$. Most of them remain to be
discovered, but most have small amplitudes \citep{rci02},
additionally diminished by the very 
frequent presence of companions
\citep{priruc2006}. Thus, the large photometric
amplitude systems discussed in our 
two papers form a tip of an iceberg which still
remains to be seen and explored.

\acknowledgements

Support from the Natural Sciences and Engineering Council of Canada
to SMR is acknowledged with gratitude.
The research made use of the SIMBAD database, operated at the CDS,
Strasbourg, France and accessible through the Canadian
Astronomy Data Centre, which is operated by the Herzberg Institute of
Astrophysics, National Research Council of Canada.
This research made also use of the Washington Double Star (WDS)
Catalog maintained at the U.S. Naval Observatory and the
General Catalog of Variable Stars maintained at the
Sternberg Astronomical Institute, Moscow, Russia.

\clearpage

\noindent
Captions to figures:

\bigskip

\figcaption[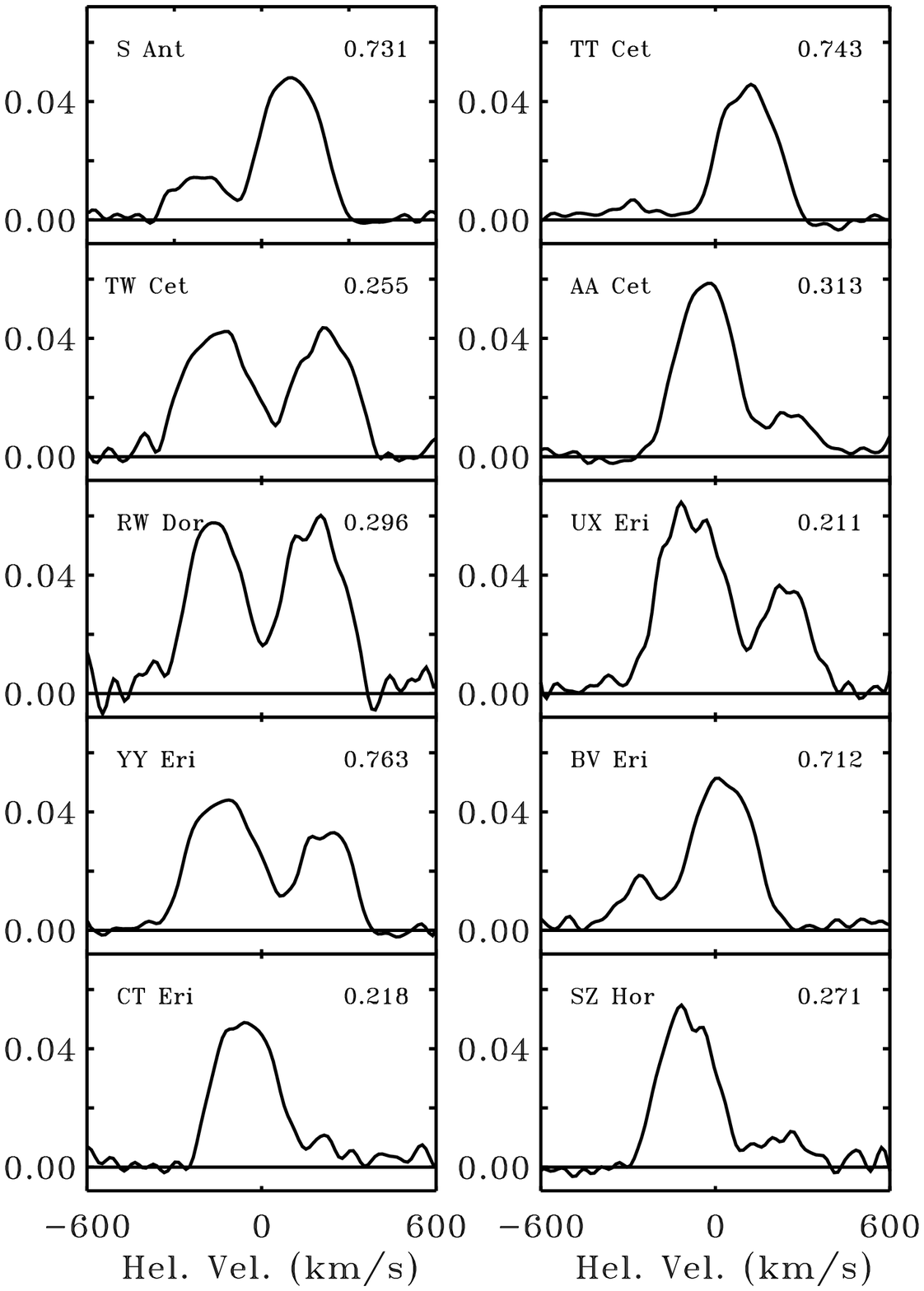]{\label{fig1} Broadening functions for 10
binary systems discussed in Section~\ref{sec2}. The orbital phase is
given in the right side of each panel. The constellation
alphabetic order in the panels has been broken by shifting the
two stars of the CU~Eri system to the next figure, to show them side
by side. Note the very weak signature of the secondary component
in TT~Cet.
}

\figcaption[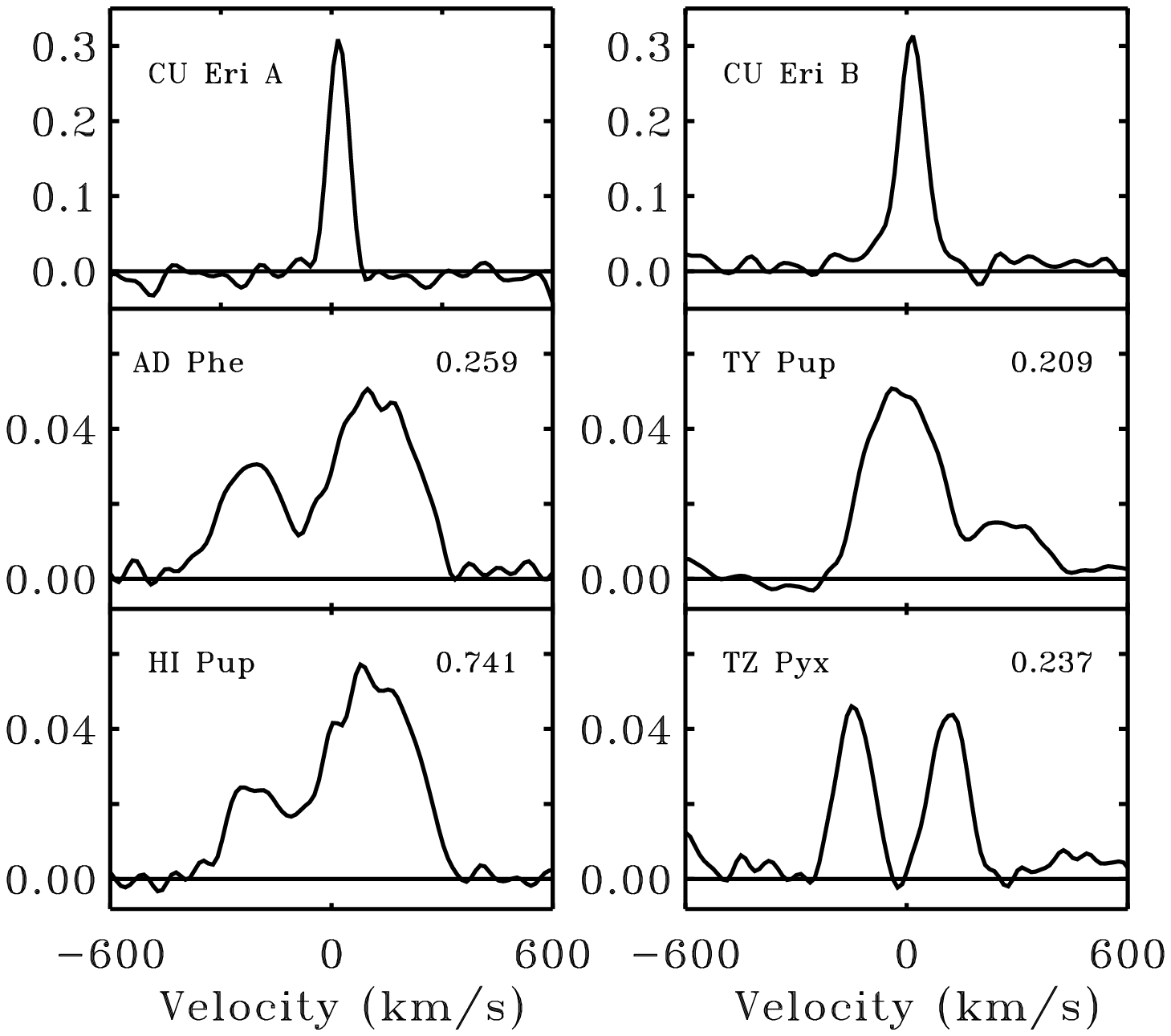]{\label{fig2} Broadening functions for the remaining
6 stars discussed in Section~\ref{sec2}. The orbital phase is
given in the right side of each panel except for the two components
of the double system CU~Eri for which photometric variability
type is unknown. We show the two components of CU~Eri to visualize a
small difference in the broadening function width and asymmetry for
CU~Eri~B. TZ~Pyx is a detached, close binary.
}

\figcaption[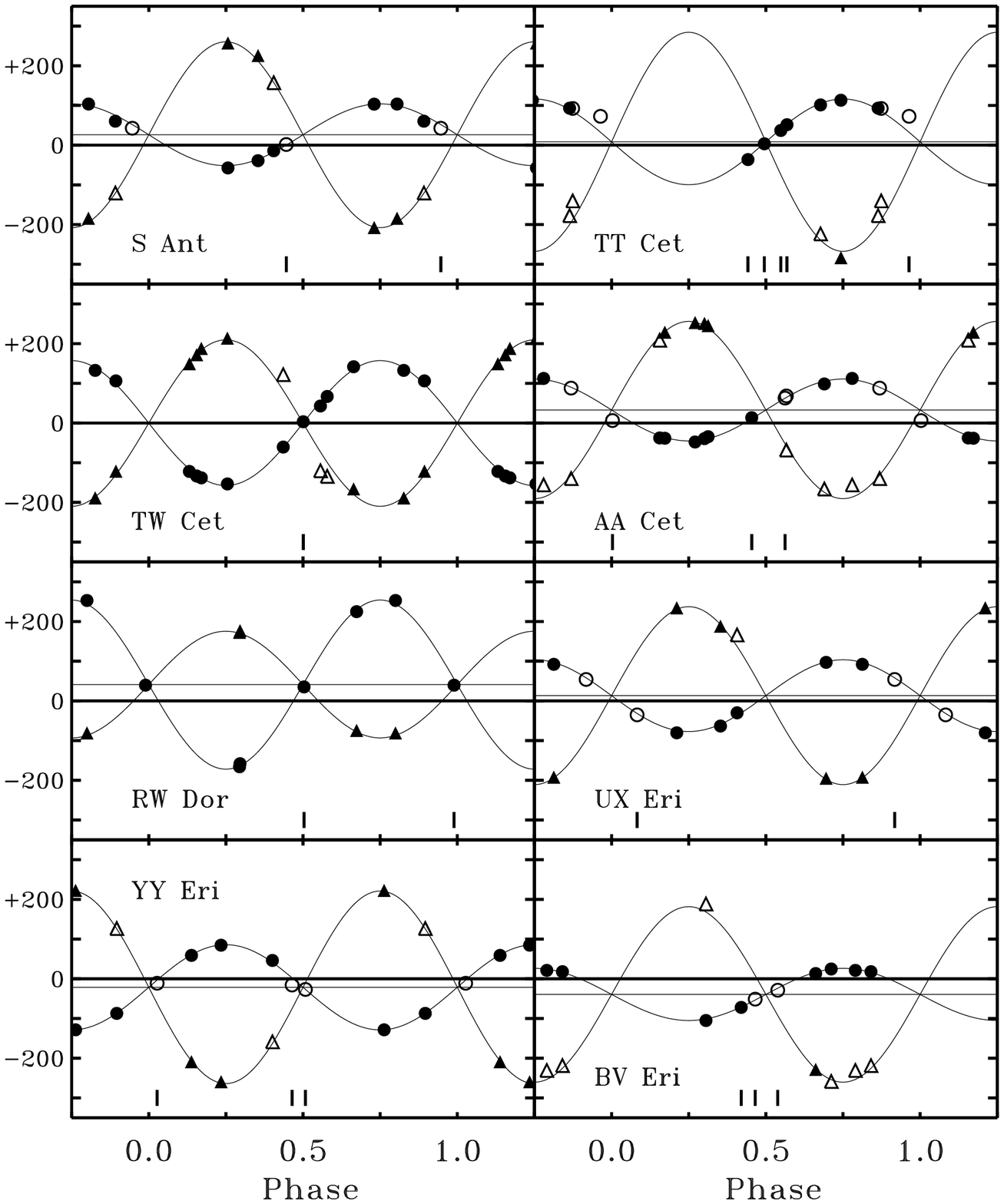]{\label{fig3} The radial velocity orbits for the
first 8 of 14 close binary systems analyzed in this paper. All systems
are SB2 binaries. RW~Dor and YY~Eri are the only W-type contact
systems while the rest are A-type systems.
The circles and triangles are used for
stronger and weaker components in the BFs while the open symbols
signify observations of lower weight in the final solutions.
The vertical dashes in the lower parts of the panels mark phases
of observations which could not be used for at least one of the
components. The thin lines give the preliminary sine curve
solutions.
}

\figcaption[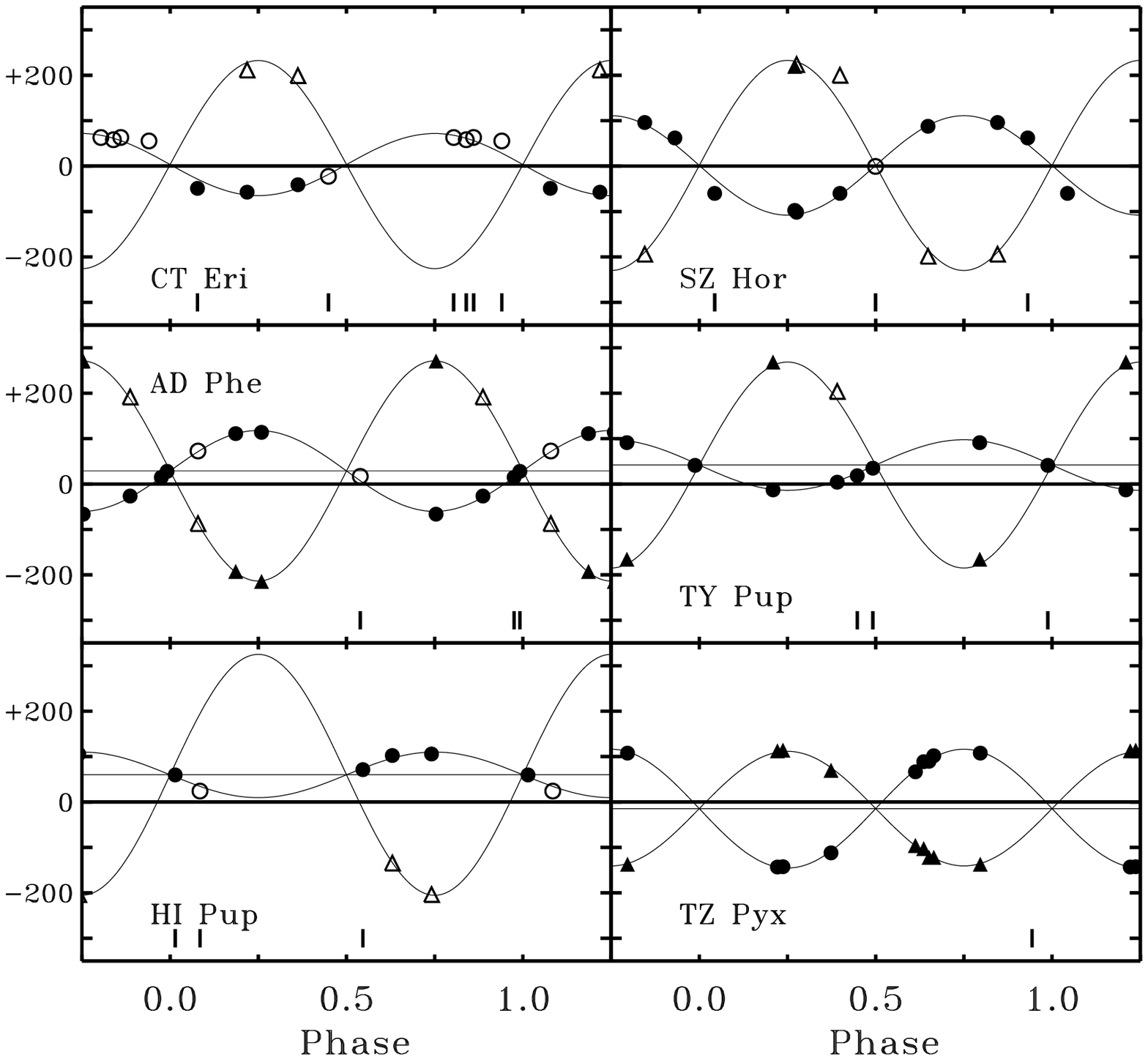]{\label{fig4} The same as in Figure~\ref{fig4}
for 6 remaining binary systems analyzed in this paper. AD~Phe
is a W-type contact binary and TZ~Pyx is a detached binary.
The remaining binaries are A-type contact systems.
}


\begin{deluxetable}{lrrrrr}


\tablewidth{0pt}
\tablenum{1}

\tablecaption{Radial velocity observations
(the first 10 observations;
the full table is available in the electronic form) \label{tab1}}
\tablehead{
\colhead{Star}	   &
\colhead{HelJD}	   &
\colhead{~$RV_1$} & \colhead{~~$W_1$} &
\colhead{~$RV_2$} & \colhead{~~$W_2$}
}
\startdata
  S Ant	  & 2450415.8015 & $ -57.33$  &	 1.0 & $ 257.22$ &  1.0 \\
  S Ant	  & 2450415.8647 & $ -39.07$  &	 1.0 & $ 225.58$ &  1.0 \\
  S Ant	  & 2450416.7571 & $ 103.22$  &	 1.0 & $-207.69$ &  1.0 \\
  S Ant	  & 2450416.8052 & $ 103.60$  &	 1.0 & $-184.36$ &  1.0 \\
  S Ant	  & 2450416.8616 & $  60.33$  &	 1.0 & $-120.54$ &  0.5 \\
  S Ant	  & 2450417.8427 & $ -14.17$  &	 1.0 & $ 157.17$ &  0.5 \\
  S Ant	  & 2450417.8691 & $   1.65$  &	 0.5 & $   0.00$ &  0.0 \\
  S Ant	  & 2450418.8422 & $  42.59$  &	 0.5 & $   0.00$ &  0.0 \\
  TT Cet  & 2450415.6540 & $  37.02$  &	 1.0 & $   0.00$ &  0.0 \\
  TT Cet  & 2450416.6000 & $   3.42$  &	 1.0 & $   0.00$ &  0.0
\enddata
\tablecomments{The table gives the radial velocities $RV_i$ and
associated weights $W_i$ for observations of binary stars described
in Section~\ref{sec2}. The velocities are expressed in km~s$^{-1}$.
The weights $W_i$ were used in the orbital solutions and can take
values of 1.0, 0.5 or 0; the zero weight observations
may be eventually used in more extensive modeling of broadening
functions.}
\end{deluxetable}

\begin{deluxetable}{lcrccccccccc}

\tabletypesize{\scriptsize}

\pagestyle{empty}

\tablecolumns{12}

\tablewidth{0pt}

\tablenum{2}
\tablecaption{Spectroscopic orbital elements \label{tab2}}
\tablehead{
   \colhead{Name} &		   
   \colhead{n$_{\rm obs}$}	&	   
   \colhead{~~~$V_0$} &		   
   \colhead{$\sigma V_0$} &	   
   \colhead{$K_1$} &		   
   \colhead{$\sigma K_1$} &	   
   \colhead{$K_2$} &		   
   \colhead{$\sigma K_2$} &	   
   \colhead{$T_0$ -- 2,400,000} &  
   \colhead{$\sigma T_0$} &	   
   \colhead{$O-C$} &		   
   \colhead{$P$}	           
}
\startdata
S Ant	 	 &     8  &  26.22  &	2.85 &	  77.84       &   3.73  &  234.10 &	 4.02	&  52,500.0659	 &  0.0049 &  $+0.0002$\tablenotemark{a} &  [0.648346]  \\[1.5ex]
TT Cet	 &     9  &   8.37  &	9.47  &	 107.9	  &  10.3   &  276    &	  27	&  52,500.1306	 &  0.0062 &  $-0.0009$ &  [0.485954]  \\[1.5ex]
TW Cet	 &    11  &  $-0.11$&	2.56 &	 157.40	  &  3.26   &  209.66 &	 3.84	&  52,500.2580	 &  0.0012 &  $-0.0029$ &  [0.316851]  \\[1.5ex]
AA Cet	 &    12  &  32.70  &	2.14 &	  77.98	  &  2.27   &  224.24 &	 4.33	&  52,500.3699	 &  0.0021 &  $+0.0047$ &  [0.536169]  \\[1.5ex]
RW Dor	 &     6  &  41.11  &	4.21 &	 134.35       &  3.46   & 213.06  &	 4.65	&  52,500.2302	 &  0.0023 &  $-0.0035$ &  [0.285463]  \\[1.5ex]
UX Eri	 &     7  &  13.27  &	2.14 &	  90.40	  &  2.64   & 224.19  &	 3.37	&  52,500.5011	 &  0.0027 &  $+0.0092$ &  [0.445286]  \\
~~~~DDO-3    &    28  &  12.79  &	1.09 &	  91.75	  &  1.55   & 245.76  &	 1.86	&  50,416.5587	 &  0.0014 &		&  [0.445279]  \\[1.5ex]
YY Eri	 &     8  & $-21.42$&	1.24 &	 107.30	  &  2.27   & 242.57  &	 2.56	&  52,500.2976	 &  0.0007 &  $-0.0098$ &  [0.321498]  \\[1.5ex]
BV Eri	 &     8  & $-39.37$&	3.26 &	  65.71	  &  2.90   & 221.1   &	 9.9	&  52,500.2700	 &  0.0044 &  $+0.0014$ &  [0.507654]  \\[1.5ex]
CT Eri	 &     8  &   3.4   &	11.06&	  68.3	  & 14.9    & 229.1   &	 48	&  52,500.2863	 &  0.0290 &  $+0.0093$ &  [0.634200]  \\[1.5ex]
SZ Hor	 &     8  &  1.43   &	5.03 &	 109.51	  & 6.60    & 231.34  &	 13.5	&  48,500.1019	 &  0.0095 &  $-0.0063$ &  [0.625118]  \\[1.5ex]
AD Phe	 &     8  &  28.87  &	1.47 &	  89.04	  & 3.10    & 242.41  &	 1.42	&  52,500.0624	 &  0.0007 &  $+0.0074$ &  [0.379923]  \\[1.5ex]
TY Pup	 &     6  &  41.86  &	4.01 &	  55.68	  & 5.52    & 226.8   & 15.4	&  52,500.1662	 &  0.0090 &  $+0.0082$ &  [0.819244]  \\[1.5ex]
HI Pup	 &     5  &  59.89  &	7.14 &	  50.2	  & 15.4   & 265.2   & 17.2    &  52,500.2462	&  0.0077 &  $-0.0108$ &  [0.432616]  \\[1.5ex]
TZ Pup	 &     9  & $-14.56$&	1.08 &	 126.22	  & 2.09    & 130.79  &	 1.52	&  52,500.3675	 &  0.0072 &  $-0.0025$ &  [2.318555]  \\
\enddata
\tablenotetext{a}{Assumes a shift of $+0.06$~d in the Cracow ephemeride, as
discussed in the text.}
\tablecomments{The column n$_{\rm obs}$ gives the number of used RV measurements.
The radial velocity parameters $V_0$, $K_1$ and $K_2$ and their rms errors are in km~s$^{-1}$.
$T_0$ is the heliocentric Julian Day of the superior conjunction (eclipse) shifted back in
time to the epoch of the original ephemeride.
The periods $P$ are in days. The fixed quantities are in square brackets.
}
\end{deluxetable}

\newpage
\plotone{f1.eps}
\newpage
\plotone{f2.eps}
\newpage
\plotone{f3.eps}
\newpage
\plotone{f4.eps}


\begin{thebibliography}{}

\bibitem[Abt(2005)]{abt05}			      
    Abt, H. A.
    2005, \apj, 629, 507

\bibitem[Baade et al.(1983)]{baade83}		      
    Baade D., Duerbeck, H. W., Karimie, M. T., \& Yamasaki, A.
    1983, \apss, 93, 69

\bibitem[Bessell(1979)]{bessell79}
    Bessell, S.M. 1979, \pasp, 91, 589

\bibitem[Chambliss(1981)]{chamb81}		      
    Chambliss, C. R.
    1981, Inf.\ Bull.\ Var.\ Stars, 2058

\bibitem[Duerbeck(1984)]{due84}			      
     Duerbeck, H. W.
     1984, \apss, 99, 363

\bibitem[ESA(1997)]{hip}				       
    European Space Agency. 1997. The Hipparcos and Tycho
    Catalogues (ESA SP-1200)(Noor\-dwijk: ESA) (HIP)

\bibitem[Gu(1999)]{gu99}			      
    Gu, S.
    1999, \aap, 346, 437

\bibitem[Gu(1993)]{gu93}			      
    Gu, S., Yang, Y., Liu, Q., \& Zhang, Z.
    1993, \apss, 203, 161


\bibitem[Hertzsprung(1928)]{hertz28}	     
    Hertzsprung, E.
    1928, Bull. Astr. Netherl., 4, 153

\bibitem[Hilditch et al.(1992)]{hild92}		 
    Hilditch, R. W., Hill, G., \& Bell, S. A.
    1992, \mnras, 255, 285

\bibitem[Hoffmeister(1956)]{hoff56}		      
    Hoffmeister, C.
    1956, Ver. St. Sonn., 3, 1

\bibitem[H\o{}g et al.(2000)]{tycho2}
    H\o{}g, E., et al. 2000, \aap, 355, L27

\bibitem[Houk(1978)]{houk78}
    Houk, N. 1978, Michigan Catalogue of 
    Two-Dimensional Spectral Types for
    the HD Stars, Vol. 2 (Ann Arbor: Univ. Michigan)

\bibitem[Houk(1982)]{houk82}
Houk, N. 1982, Michigan Catalogue of 
    Two-Dimensional Spectral Types for
    the HD Stars, Vol. 3 (Ann Arbor: Univ. Michigan)

\bibitem[Houk \& Cowley(1975)]{houketal75}
Houk, N., \& Cowley, A. P. 1975, Michigan Catalogue of 
    Two-Dimensional Spectral Types for
    the HD Stars, Vol. 1 (Ann Arbor: Univ. Michigan)

\bibitem[Houk \& Smith-Moore(1988)]{houketal88}
Houk, N., \& Smith-Moore, M. 1988, Michigan Catalogue of 
    Two-Dimensional Spectral Types for
    the HD Stars, Vol. 4 (Ann Arbor: Univ. Michigan)

\bibitem[Houk \& Swift(1999)]{houketal99}
Houk, N., \& Swift, C. 1999, Michigan Catalogue of 
    Two-Dimensional Spectral Types for
    the HD Stars, Vol. 5 (Ann Arbor: Univ. Michigan)


\bibitem[H{\o}g et~al.(2000)]{Tycho2}		      
    H{\o}g, E., Fabricius, C., Makarov, V. V., Urban, S., Corbin, T.,
    Wycoff, G., Bastian, U., \\
    Schwekendiek, P., \& Wicenec, A.
    2000, \aap, 355L, 27 (Tycho-2)

\bibitem[Huruhata et al.(1953)]{huruhata53}
    Huruhata M., Dambara, Y., \& Kitamura, M.
    1953, Tokio Ann.\ Sec.\ Ser., 3, 227

\bibitem[Ka{\l}u\.{z}ny(1986)]{kal86}
    Ka{\l}u\.{z}ny, J.
    1986, \pasp, 98, 662

\bibitem[Kreiner(2004)]{kreiner2004}                   
    Kreiner, J.M.
    2004, Acta Astron., 54, 207

\bibitem[Kreiner et al.(2001)]{kreiner2001}
    Kreiner, J.M., Kim, C.H., Nha, I.S.
    2001, ``An atlas of (O-C) diagrams of eclipsing binary stars'',
    Wydawnictwo Naukowe Akademii Pedagogicznej, Krakow

\bibitem[Maceroni et al.(1982)]{macer82}	      
    Maceroni, C., Milano, L., \& Russo, G.
    1982, \apss, 49, 123

\bibitem[Marton et al.(1989)]{mart89}
    Marton S. F., Grieco, A., \& Sistero, R. F.
    1989, \mnras, 240, 931

\bibitem[McFarlane \& Hilditch(1987)]{McFH87}	      
    McFarlane, T. M. \& Hilditch, R. W.
    1987, \mnras, 227, 381

\bibitem[Nesci et al.(1986)]{nesci86}		      
    Nesci, R., Maceroni, C., Milano, L., \& Russo, G.
    1986, \aap, 159, 142.

\bibitem[Nordstrom et al.(2004)]{nord04}
    Nordstr\o{}m, B., et al. 2004, \aap, 418, 989	

\bibitem[Popper(1956)]{popper56}
    Popper, D. M.
    1956, \apj, 124, 208

\bibitem[Pribulla \& Rucinski(2006)]{priruc2006}       
    Pribulla, T., Rucinski, S.M.
    2006, \aj, 131, 2986

\bibitem[Pribulla et~al.(2003)]{pri03}
    Pribulla, T., Kreiner, J. M., \& Tremko, J.
    2003, Contrib. Astron. Obs. Skalnat\'e Pleso, 33, 38

\bibitem[Pribulla et~al.(2006)]{ddo11}			
    Pribulla, T., Rucinski, S. M., Lu, W., Mochnacki, S. W.,
    Conidis, G., R., Blake, R. M., DeBond, H., Thomson, J. R.,
    Pych, W., Ogloza, W., \& Siwak, M.
    2006, \aj, 132, 769

\bibitem[Rucinski(2001)]{rci01}			
    Rucinski, S. M.
    2001, \aj, 122, 1007

\bibitem[Rucinski(2002)]{rci02}			
    Rucinski, S. M.
    2002, \pasp, 114, 1124

\bibitem[Rucinski(2006)]{rci06}			
    Rucinski, S. M.
    2006, \mnras, 368, 1319

\bibitem[Rucinski \& Duerbeck(2006)]{RD06}		
    Rucinski, S. M., \& Duerbeck, H.
    2006, \aj, 132, 1539

\bibitem[Rucinski, Lu, \& Mochnacki(2000)]{ddo3}	
    Rucinski, S. M., Lu, W., \& Mochnacki, S. W.
    2000, \aj, 120, 1133 (DDO-3)

\bibitem[Russo et al.(1982)]{russo82}		      
    Russo, G., Sollazzo, C., Maceroni, C., \& Milano, L.
    1982, \aaps, 47, 211

\bibitem[Sahade \& Ber\'{o}n D\`{a}vila(1963)]{sahade63}	 
    Sahade, J. \& Ber\'{o}n D\`{a}vila, F.
    1963 Ann. d'Ap., 26, 153

\bibitem[Shaw(1994)]{shaw94}			      
    Shaw, J. S.
    1994, Mem.\ Soc.\ Astr.\ Ital., 65, 95

\bibitem[Srivastava \& Srivastava(1987)]{ss87}	      
    Srivastava, J. B. \& Srivastava, R. K.
    1987, \apss, 129, 415

\bibitem[Strohmeier(1966)]{stroh66}		      
    Strohmeier, W.
    1966, Inf.\ Bull.\ Var.\ Stars, 158

\bibitem[Strohmeier(1967a)]{stroh67a}			
    Strohmeier, W.
    1967a, Inf.\ Bull.\ Var.\ Stars, 199

\bibitem[Strohmeier(1967b)]{stroh67b}		      
    Strohmeier, W.
    1967b, Inf.\ Bull.\ Var.\ Stars, 225

\bibitem[Strohmeier(1968)]{stroh68}		      
    Strohmeier, W.
    1968, Inf.\ Bull.\ Var.\ Stars, 262

\bibitem[Strohmeier \& Bauernfeind(1969)]{stroh69}    
    Strohmeier, W. \& Bauernfeind, H.
    1969, Inf.\ Bull.\ Var.\ Stars, 360

\bibitem[Struve(1947)]{struve47}		      
    Struve, O.
    1947, \apj, 106, 92

\bibitem[Struve(1950)]{struve50a}		      
    Struve, O.
    1950, \apj, 112, 184

\bibitem[Struve et al.(1950)]{struve50}		      
    Struve, O., Horak, H. G., Canavaggia, R.,
    Kourganoff, V., \& Colacevich, A.
    1950, \apj, 111, 658

\bibitem[Wolf \& Kern(1983)]{WK83}		      
   Wolf, G. W. \& Kern, J. T.
    1983, \apjs, 52, 429

\bibitem[Wood et al.(1980)]{wood}		      
   Wood, F.B., Oliver, J.P., Florkowski, D.R. \& Koch, R.H.
   1980, A finding list for Observers of 
   Interacting Binary Stars, University
   of Pennsylvania Press, Philadelphia

\end{thebibliography}
\end{document}